\documentclass{article}
\pdfoutput=1

\usepackage{jheppub}
\usepackage{amsmath}
\usepackage{amssymb}
\usepackage{booktabs}
\usepackage{braket}
\usepackage{colortbl} 
\usepackage{comment}
\usepackage{dcolumn}
\usepackage{graphicx}
\usepackage[utf8]{inputenc}
\usepackage{multirow}
\usepackage{siunitx}
\usepackage[normalem]{ulem}
\usepackage{xcolor}
\usepackage{xspace}
\usepackage{hyperref}
\usepackage[nameinlink]{cleveref}

\newcommand{\ie}{\textit{i.e.}\xspace}

\newcommand{\GeV}{\textrm{GeV}\xspace}

\newcommand{\GFermi}{\ensuremath{G_\textrm{F}}\xspace}
\newcommand{\dynesty}{\texttt{dynesty}\xspace}
\newcommand{\EOS}{\texttt{EOS}\xspace}

\newcommand{\tildeVcs}{\tilde{V}_{cs}}
\newcommand{\absVcs}{\ensuremath{|V_{cs}|}\xspace}
\newcommand{\wc}[2][\ell]{\ensuremath{\mathcal{C}^{#1}_{#2}}\xspace}
\newcommand{\vecnu}{\ensuremath{\vec{\nu}}\xspace}
\newcommand{\vecth}{\ensuremath{\vec{\vartheta}}\xspace}
\renewcommand{\Im}{\operatorname{Im}}
\renewcommand{\Re}{\operatorname{Re}}

\begin{document}

\title{Constraining $\boldsymbol{|V_{cs}|}$ and physics beyond the Standard Model from exclusive (semi)leptonic charm decays}
\author[a,b]{Carolina Bolognani,}
\emailAdd{carolina.bolognani@cern.ch}
\author[c]{Méril Reboud,}
\emailAdd{meril.reboud@cnrs.fr}
\author[d]{Danny van Dyk,}
\emailAdd{danny.van.dyk@gmail.com}
\author[a,b]{K. Keri Vos}
\emailAdd{k.vos@maastrichtuniversity.nl}

\affiliation[a]{Gravitational Waves and Fundamental Physics (GWFP), Maastricht University, Duboisdomein 30, NL-6229 GT Maastricht, the Netherlands}
\affiliation[b]{Nikhef, Science Park 105, NL-1098 XG Amsterdam, the Netherlands}
\affiliation[c]{Université Paris-Saclay, CNRS/IN2P3, IJCLab, 91405 Orsay, France}
\affiliation[d]{Institute for Particle Physics Phenomenology and Department of Physics, Durham University, Durham DH1 3LE, UK}

\abstract{
We study the available data on exclusive leptonic and semileptonic $c\to s\ell^+\nu$ decays within the Standard Model and beyond.
Our analysis accounts for theory correlations between the relevant hadronic matrix elements through application of dispersive bounds.
We find that, within a global analysis, the dispersive bounds are generally well respected and only mildly affect
the extraction of the Cabibbo-Kobayashi-Maskawa (CKM) matrix element $|V_{cs}|$.
Assuming Standard Model dynamics, we obtain
$$
    |V_{cs}| = 0.957 \pm 0.003\,,
$$
which is compatible with the HFLAV/PDG reference value $\absVcs = 0.975 \pm 0.006$ at the $2.7\,\sigma$ level.
Our findings lead to significant deficits in the second-row and second-column unitarity relations of the CKM matrix.
Allowing for beyond the SM contributions in the Weak Effective Theory, we find very strong constraints on potential (pseudo)scalar and tensor effects.
However, the data still permits sizeable CP-violating right-handed currents.
}

\begin{flushright}
    EOS-2024-02\\
    IPPP/24/36\\
    Nikhef-2024-11\\
\end{flushright}
\vspace*{-3\baselineskip}

\maketitle

\section{Introduction}
\label{sec:intro}

The Cabibbo-Kobayashi-Maskawa (CKM) quark mixing matrix is a central parameter in the Standard Model of particle physics (SM).
Precision determinations of its elements are key to understanding the origin of (quark) flavour.
The CKM element $|V_{cs}|$ is currently the second-most precisely known CKM element.
The Particle Data Group (PDG) has determined a reference value for $\absVcs$ with a relative uncertainty
of $0.6\%$~\cite{PDGCKMReview}.
In computing this value, the contemporary world averaged of the total branching ratios of $D\to \bar{K}\ell^+\nu$ and $D_s^+\to \ell^+\nu$ decays are used
in combination with theoretical determinations of the relevant hadronic form factors and decay constants.
\\

In this work, we study exclusive flavour-changing $c\to s\ell^+\nu$ processes in a global fit.
These processes include $\bar{D}_s\to \ell^+\nu$, $\bar{D}_s^*\to \ell^+\nu$, $D\to \bar{K}\ell^+\nu$,
and $\Lambda_c\to\Lambda\ell^+\nu$.
%
First, we determine if the available data can be simultaneously described by the available theory predictions.
This question is not trivial since the hadronic matrix elements are connected within the framework of dispersive bounds~\cite{Okubo:1973tj,Boyd:1997kz,Caprini:1997mu}.\\

We determine $\absVcs$ from the global fit within this setup and discuss the agreement of the result with CKM unitarity.
We also probe if determinations of $\absVcs$ from the individual
decay processes agree with each other and with the quoted PDG value~\cite{PDGCKMReview}.
Finally, we determine how Beyond the Standard Model (BSM) effects are constrained by
$c\to s\ell^+\nu$ processes by determining the maximally allowed parameter
space within the $sc\nu\ell$ sector of the Weak Effective Theory (WET)~\cite{Aebischer:2017gaw,Jenkins:2017jig,Jenkins:2017dyc}.
Our analysis leverages procedures and techniques first developed in the context of
a study of $b\to u\ell^-\bar\nu$ decays~\cite{Leljak:2023gna}.

Two previous studies of leptonic and semileptonic $c\to s\ell^+\nu$ decays are available in
Refs.~\cite{Fajfer:2015ixa,Becirevic:2020rzi}. Our study improves upon both analyses in three ways.
First, we consider two additional decays, $D_s^{*+}\to e^+\nu$ and $\Lambda_c\to \Lambda\ell^+\nu$,
which both provide complementary information compared to the (semi)leptonic
decays investigated in these works.
Second, we include updates to the measurements used in these works, primarily by the BESIII experiment.
Third, we use new lattice QCD results for the required hadronic matrix elements.
Moreover, we account for the first time for the dispersive bound connecting the various hadronic matrix elements in exclusive $c\to s\ell^+\nu$ processes.
In this, we closely follow what has been done in Ref.~\cite{Gubernari:2023puw} for
local form factors in rare $b$ decays.\\
In addition, our study is the first to simultaneously account for the full set of Wilson coefficients of a semileptonic sector of the WET
and all hadronic nuisance parameters using dispersive bounds in a joint analysis.
For this analysis, we strictly assume lepton-flavour universality.

Further studies include Refs.~\cite{Fleischer:2019wlx,Leng:2020fei}, which focus on investigating the potential for lepton-flavour universality violation in $c\to s\ell^+\nu$
and Ref.~\cite{Mahata:2024kxu}, which focuses on exploring $c\to s\ell^+\nu$
physics through the exclusive decays $D_s^+\to \eta^{(\prime)} \ell^+\nu$.
\\

The structure of this paper is as follows. We discuss the analysis setup in \cref{sec:setup},
describing the theoretical framework (in \cref{sec:setup:framework}), the statistical approach (in \cref{sec:setup:approach}),
the experimental data used (in \cref{sec:setup:exp}), our choice of statistical models and parameters of interest
(in \cref{sec:setup:models}), and our choice of prior for the hadronic nuisance parameters (in \cref{sec:setup:nuisances}).
\Cref{sec:results} is dedicated to documenting the methods used before presenting our
numerical results, with subsections dedicated to the main objectives.
We discuss the compatibility of theory predictions and measurements in \cref{sec:results:SM},
our determinations of \absVcs in \cref{sec:results:CKM}, the implications for CKM unitarity in
\cref{sec:results:CKM-unitarity}, and constraints on potential BSM effects in \cref{sec:results:BSM}.
We conclude in \cref{sec:conclusion}. Our treatment of the hadronic matrix elements is
documented in detail in \cref{app:hme}.

\section{Analysis Setup}
\label{sec:setup}

\subsection{Theoretical Framework}
\label{sec:setup:framework}

The dimension-six effective Hamiltonian of the $sc\nu\ell$ sector can be normalised as~\cite{Aebischer:2017gaw,Jenkins:2017jig,Jenkins:2017dyc}\footnote{%
    Here ``sector'' refers to a set of operators in the Hamiltonian that
    do not mix with other terms at leading-order in $\GFermi \sim g^2/M_W^2$~\cite{Aebischer:2017ugx}.
}
\begin{equation}
    \label{eq:intro:Heff}
    \mathcal{H}^{sc\nu\ell}
        = -\frac{4\GFermi}{\sqrt{2}} \tildeVcs^* \sum_i \wc{i}(\mu_c) \mathcal{O}_i^\ell + \text{h.c.}\,,
\end{equation}
where $\wc{i}$ are the Wilson coefficients and $\mathcal{O}_i$ the local field operators.
The normalisation in terms of the Fermi constant $\GFermi$ and an arbitrary constant $\tildeVcs$
proves to be convenient later on.
The Wilson coefficients encode the dynamics of the full theory, either the SM or any viable BSM theory, above
the separation scale $\mu_c \simeq 1.275\,\GeV$. The matrix elements of the operators encode the dynamics below
the separation scale. Here, we assume that there are no right-handed neutrinos of mass less than $\mu_c$.
Consequently, the basis of operators of mass dimension six reduces to five independent operators. Our
choice of basis reads
\begin{equation}
\begin{aligned}
    \mathcal{O}_{V,L}^\ell
        & = [\bar{s} \gamma^\mu  P_L c]\, [\bar{\nu} \gamma_\mu      P_L \ell]\,,    &
    \mathcal{O}_{V,R}^\ell
        & = [\bar{s} \gamma^\mu  P_R c]\, [\bar{\nu} \gamma_\mu      P_L \ell]\,,    \\
    \mathcal{O}_{S,L}^\ell
        & = [\bar{s}             P_L c]\, [\bar{\nu}                 P_L \ell]\,,    &
    \mathcal{O}_{S,R}^\ell
        & = [\bar{s}             P_R c]\, [\bar{\nu}                 P_L \ell]\,,    \\
    \mathcal{O}_{T}^\ell
        & = [\bar{s} \sigma^{\mu\nu} b]\, [\bar{\nu} \sigma_{\mu\nu} P_L \ell]\,.
\end{aligned}
\end{equation}
Matching the effective Lagrangian to the SM amplitudes, one finds that~\cite{Sirlin:1980nh}
\begin{equation}
    \label{eq:setup:WET:WCsir}
    \wc{V,L}(\mu) = 1 + \frac{\alpha_e}{\pi} \ln \left(\frac{M_Z}{\mu}\right)\, \simeq 1.01,
\end{equation}
while all other Wilson coefficients are zero. In the SM, we identify $\tildeVcs^*$ with the
conjugated CKM matrix element $V_{cs}^*$. Beyond the SM, the Wilson coefficients provide a low-energy
footprint of the genuine BSM dynamics at or above the electroweak scale.
These coefficients can then be used to constrain the parameters of a UV-complete BSM model.\\

\subsection{Approach}
\label{sec:setup:approach}

Our analysis follows the Bayesian approach to statistics, focusing on minimising or sampling from the posterior probability density function (PDF)
\begin{equation}
    \label{eq:setup:posterior}
    P(\vecth, \vecnu \,|\, D, M) \propto P(D \,|\, \vecth, \vecnu, M)\, P_0(\vecth, \vecnu \,|\, M)\,.
\end{equation}
Here, $\vecth$ represents the parameters of interest, and $\vecnu$ represents the nuisance parameters.
The latter arise exclusively from our description of hadronic matrix elements in the observables.
The posterior PDF is proportional to the likelihood function $P(D \,|\, \vecth, \vec\nu, M)$, which encapsulates
the experimental and theoretical constraints imposed by the data $D$ under consideration, and to the prior PDF
$P_0(\vecnu, \vecth \,|\, M)$, which accounts for our prior knowledge about the parameters in the model $M$.
We discuss the various datasets and fit models entering our analysis in \cref{sec:setup:exp} and \cref{sec:setup:models}, respectively.\\

To compare two models $M_1$ and $M_2$, it is instrumental to determine the normalisation
of \cref{eq:setup:posterior} for either model and a common dataset $D$, i.e.,
\begin{equation}
    \label{eq:setup:evidence}
    P(D \,|\, M_i) \equiv \iint d\vecth\,d\vecnu\, P(D \,|\, \vecth, \vecnu, M_i)\, P_0(\vecth, \vecnu \,|\, M_i)\,.
\end{equation}
Their ratio, the so-called Bayes factor $K \equiv P(D \,|\, M_1) / P(D \,|\, M_2)$, then provides information
about the efficiency of the two models in describing the common dataset $D$.
Following Jeffreys' interpretation~\cite{Jeffreys:1939xee}, the model $M_1$ is preferred over the model $M_2$ if $K > 1$.
This preference can be characterised either as strong if $10 < K < 100$ or as decisive if $100 < K$.
A Bayes factor $3 < K < 10$ is still interpreted to be substantially in favour of $M_1$ over $M_2$,
whereas a Bayes factor of $1 < K < 3$ is ``barely worth mentioning''.
For $K < 1$, the above interpretation favours $M_2$ over $M_1$ with the replacement $K \to 1/K$.

\subsection{Experimental Data}
\label{sec:setup:exp}

Our analysis uses the following experimental data of exclusive $c\to s\ell^+\nu$ processes:
\begin{description}
    \item[$\boldsymbol{D_s^+\to\lbrace \mu^+, \tau^+\rbrace\nu}$]
    For $D_s^+\to \mu^+\nu$, the BESIII experiment has contributed a recent measurement~\cite{BESIII:2023cym} that is not yet included in the world averages compiled by the HFLAV~\cite{Amhis:2022vcd} and PDG~\cite{ParticleDataGroup:2022pth} collaborations.
    Using this new result together with results by the BaBar~\cite{BaBar:2010ixw}, Belle~\cite{Belle:2013isi}, BESIII~\cite{BESIII:2016cws,BESIII:2021anh}, and CLEO-c~\cite{CLEO:2009lvj} experiments, we obtain as the world average as of 2024
    \begin{equation}
        \mathcal{B}(D_s^+ \to \mu^+\nu)_\text{avg}=(5.35 \pm 0.10) \times 10^{-3}\,.
    \end{equation}
    For $D_s^+\to \tau^+\nu$, the situation is similar,
    with BESIII measurements~\cite{BESIII:2023ukh,BESIII:2023fhe} appearing after the most recent HFLAV/PDG
    world average, which itself is based on Refs.~\cite{CLEO:2009lvj,CLEO:2009jky,CLEO:2009vke,BaBar:2010ixw,Belle:2013isi,BESIII:2016cws,BESIII:2021anh}.
    We obtain the world average as of 2024
    \begin{equation}
        \mathcal{B}(D_s^+\to \tau^+\nu)_\text{avg}=(5.39 \pm 0.12)\%\,.
    \end{equation}

    These measurements contribute a total of $2$ observations for our analysis.

    \item[$\boldsymbol{D_s^{*+} \to e^+\nu}$] For $D_s^{*+} \to e^+\nu$, we use a recent measurement of the BESIII experiment~\cite{BESIII:2023zjq}
    \begin{equation}
        \mathcal{B}(D_s^{*+} \to e^+\nu)
          = (2.1^{+1.2}_{-0.9}\pm 0.2) \times 10^{-5}\,.
    \end{equation}
    However, the total decay width of the $D_s^{*+}$ is currently unknown from any experiment.
    Instead, we use the prediction for the decay width determined in Ref.~\cite{Meng:2024gpd}.
    This value is obtained from the PDG world average of the experimental measurements of the dominant branching ratio
    $\mathcal{B}(D_s^{*+} \to D_s^+ \gamma)$~\cite{ParticleDataGroup:2022pth} and a lattice QCD prediction for the partial decay
    width $\Gamma(D_s^{*+} \to D_s^+ \gamma)$~\cite{Meng:2024gpd}.   

    This measurement contributes a total of $1$ observation for our analysis.

    \item[$\boldsymbol{D^0\to K^-\lbrace e^+, \mu^+ \rbrace\nu}$]
    For $D^0\to K^-e^+\nu$, we use the PDG world average of
    the branching ratio measurement~\cite{ParticleDataGroup:2022pth}
    \begin{equation}
        \mathcal{B}(D^0 \to K^-e^+\nu) = (3.525 \pm 0.023)\%\,,
    \end{equation}
    which is based on results by the Belle~\cite{Belle:2006idb}, BES~\cite{BES:2004rav}, BESIII~\cite{BESIII:2015tql,BESIII:2021mfl}, and CLEO-c~\cite{CLEO:2009svp} experiments.
    This average is dominated by the BESIII measurement in Ref.~\cite{BESIII:2015tql}.
    
    For $D^0\to K^-\mu^+\nu$, we use the world average
    of the branching ratio measurements~\cite{ParticleDataGroup:2022pth}
    by the Belle~\cite{Belle:2006idb} and BESIII~\cite{BESIII:2018ccy} experiments
    \begin{equation}
        \mathcal{B}(D^0 \to K^-\mu^+\nu) = (3.41 \pm 0.04)\%\,,
    \end{equation}
    which is dominated by the BESIII measurement
    in Ref.~\cite{BESIII:2018ccy}.
    In addition, we use the available $q^2$-binned differential rate for both decays from Refs.~\cite{BESIII:2015tql,BESIII:2018ccy},
    which we convert to the normalised decay rate
    $1/\Gamma\, d\Gamma/dq^2$.\\

    These measurements contribute a total of $36$ observations for our analysis.

    \item[$\boldsymbol{D^+\to \bar{K}^0\lbrace e^+, \mu^+ \rbrace\nu}$] For $D^+\to \overline{K}^0 e^+\nu$, we use the world average \cite{ParticleDataGroup:2022pth},
    \begin{equation}
        \mathcal{B}(D^+\to \bar{K}^0 e^+\nu) = (8.72 \pm 0.09)\%\,,
    \end{equation}
    based on measurements from BES~\cite{BES:2004obp}, BESIII~\cite{BESIII:2015jmz,BESIII:2016hko,BESIII:2017ylw,BESIII:2021mfl}.
    The differential decay rate is also available from Ref.~\cite{BESIII:2017ylw}, which we implement as the normalised decay rate $1/\Gamma\, d\Gamma/dq^2$. 
    For $D^+\to \bar{K}^0 \mu^+\nu$, we use the only available results coming from BESIII~\cite{BESIII:2016gbw},
    \begin{equation}
        \mathcal{B}(D^+\to \bar{K}^0 \mu^+\nu) = (8.72 \pm 0.07 \pm 0.18)\%\,.
    \end{equation}

    These measurements contribute a total of $10$ observations for our analysis.
    
    \item[$\boldsymbol{\Lambda_c\to \Lambda\ell^+\nu}$]
    For $\Lambda_c\to \Lambda \lbrace e^+,\mu^+\rbrace \nu$ decays, we use measurements of the branching ratios by
    the BESIII experiment~\cite{BESIII:2022ysa,BESIII:2023vfi}, yielding
    \begin{equation}
    \begin{aligned}
        \mathcal{B}(\Lambda_c\to \Lambda e^+\nu)   & = (3.56 \pm 0.11 \pm 0.07)\%\,, \\
        \mathcal{B}(\Lambda_c\to \Lambda \mu^+\nu) & = (3.48 \pm 0.14 \pm 0.10)\%\,.
    \end{aligned}
    \end{equation}
    Although measurements of the differential distributions for the decay chain $\Lambda_c\to \Lambda(\to p \pi) e^+\nu$
    have been undertaken by BESIII~\cite{BESIII:2022ysa,BESIII:2023vfi}, the data have not been made public.
    The differential data strongly constrain the BSM parameters space, as discussed in the context of
    $\Lambda_b\to \Lambda_c(\to \Lambda \pi) \ell^-\bar\nu$~\cite{Boer:2019zmp}; the analogue of the $\Lambda_c$
    decays among semileptonic $b\to c\ell^-\bar\nu$ processes.
    BESIII also recently measured for the first time the leptonic and hadronic asymmetries of $\Lambda_c\to \Lambda \ell\nu$ decays~\cite{BESIII:2023vfi}.
    However, we cannot use these measurements in our global analysis, because of the absence of publicly available information on their correlations.\\

    These measurements contribute a total of $2$ observations for our analysis.
\end{description}
We thus have a total of $51$ observations.\\

We do not use decays to $\eta^{(\prime)}$ because their form factors, while available from light-cone sum rules and lattice QCD~\cite{Duplancic:2015zna,Hu:2021zmy,Bali:2014pva},
are not yet as stringently constrained as the ones for the other modes used in this analysis.
We also do not use measurements of $D^+\to \bar{K}^{*0}\ell^+\nu$ nor $D_s^+\to \phi \ell^+\nu$, because their hadronic final states are unstable vector resonances.
As such, they may suffer from $S$-wave pollution, which is a problem from the point of view of both the experimental extraction of the observables and the theoretical predictions.
Therefore, we do not consider them to be competitive with the processes included above.

\subsection{Parameters of Interest and their Priors}
\label{sec:setup:models}

We analyse the available data using three fit models, which we label SM, CKM, and WET. These models share a common set
of (hadronic) nuisance parameters $\vecnu$ but differ in terms of the parameters of interest $\vecth$.

\begin{description}
    \item[\textbf{SM}] This model has no parameters of interest. We use a fixed normalisation $\tildeVcs = 0.975$, corresponding to the average determined by the Particle Data Group~\cite{PDGCKMReview}.
    The left-handed Wilson coefficient $\wc{V,L}$ is set to its SM value in \eqref{eq:setup:WET:WCsir}, and all other Wilson coefficients are fixed to zero.
    This fit model serves as the \emph{null hypothesis} for our model comparisons.

    \item[\textbf{CKM}] This model has a single parameter of interest: $|\tildeVcs|$,
    which is floated within the interval $[0.88, 1.03]$ with a uniform prior PDF.
    The Wilson coefficients are treated as in the SM fit model.

    \item[\textbf{WET}] This model has nine parameters of interest, which describe the degrees of freedom
    for the five complex Wilson coefficients that appear in the effective Hamiltonian \cref{eq:intro:Heff}.
    The tenth degree of freedom, which can be chosen by imposing $\arg \tildeVcs = 0 = \arg{\wc{V,L}}$,
    is fixed since the overall phase of the effective Hamiltonian is unobservable.
    To provide an absolute scale for the magnitude of these Wilson
    coefficients, we fix $|\tildeVcs| = 0.975$, as in the SM fit model.
    We assume lepton-flavour universality of the Wilson coefficients.\\
    The parameters are floated in intervals chosen to fully contain
    the likelihood, using independent uniform prior PDFs.
    These intervals are
    \begin{equation}
    \begin{aligned}
         0.88  & \leq \Re\wc{V,L} \leq  1.03 \,,   &
         \\
        -0.05  & \leq \Re\wc{V,R} \leq +0.02 \,,   &
        -0.8   & \leq \Im\wc{V,R} \leq +0.8  \,,   \\
        -0.055 & \leq \Re\wc{S,L} \leq +0.055\,,   &
        -0.1   & \leq \Im\wc{S,L} \leq +0.1  \,,   \\
        -0.07  & \leq \Re\wc{S,R} \leq +0.055\,,   &
        -0.1   & \leq \Im\wc{S,R} \leq +0.1  \,,   \\
        -0.12  & \leq \Re\wc{T}   \leq +0.12 \,,   &
        -0.25  & \leq \Im\wc{T}   \leq +0.25 \,,
    \end{aligned}
    \end{equation}
\end{description}

\subsection{Hadronic Nuisance Parameters and their Priors}
\label{sec:setup:nuisances}

At leading order in $\alpha_e$, the leptonic and hadronic matrix elements factorise.
Non-factorisable corrections occur only at $\mathcal{O}(\alpha_e) \simeq 1/137$.
Hard virtual corrections are included as part of the SM value for the Wilson coefficient $\wc{V,L}$~\cite{Sirlin:1980nh}
while soft real radiation and soft structure-independent corrections are accounted for in the experimental data using the PHOTOS software~\cite{Golonka:2005pn}.
Structure-dependent non-factorisable corrections are currently not available for exclusive $c\to s\ell^+\nu$ decays.\\

In our analysis, hadronic matrix elements are described by a substantial number of nuisance parameters.
They need to be varied to account for the theoretical uncertainties of the hadronic matrix elements.
In the case of leptonic decays, the nuisance parameters are decay constants of the decaying hadrons.
In the case of semileptonic decays, the nuisance parameters relate to hadronic form factors, \ie real-valued functions of $q^2 = m_{\ell^+\nu}^2$. 
The semileptonic nuisance parameters depend on our choice of parametrisation for these form factors.
For this analysis, we employ a form factor parametrisation that respects dispersive bounds,
leading to controlled (\ie, parametric) systematic uncertainties in the fit.
A summary of the hadronic nuisance parameters pertaining to this analysis is provided below.
We refer to \cref{app:hme} for further details on the definitions of these parameters.

\begin{description}
    \item[$\boldsymbol{D_s^+\to \ell^+\nu}$] Leptonic decays of the pseudoscalar $D_s^+$ meson are described by single decay constant $f_{D_s}$; see \cref{eq:hme:defs:psd-decay-const} for the definition.
    This quantity is well known from lattice QCD studies. We use the world average~\cite{FlavourLatticeAveragingGroupFLAG:2021npn}
    of $N_f = 2 + 1 + 1$ results by the ETM~\cite{Carrasco:2014poa} and FNAL/MILC~\cite{Bazavov:2017lyh} collaborations
    \begin{equation}
        f_{D_s} = 0.2499 \pm 0.0005\,\GeV\,,
    \end{equation}
    at a precision of 0.2\%. The total number of nuisance parameters is $1$.
    
    \item[$\boldsymbol{D_s^{*+}\to \ell^+\nu}$] Leptonic decays of the pseudoscalar $D_s^+$ meson are described by a vector decay constants $f_{D_s^*}$ and a tensor decay constant $f_{D_s^*}^T$;
    see \cref{eq:hme:defs:vec-decay-consts} for their definitions.
    The vector $D_s^{*+}$ decay constant is not as well known as the pseudoscalar $D_s^+$ decay constant.
    It has been determined from a recent lattice QCD analysis~\cite{Meng:2024gpd} with $\sim 2\%$ precision
    \begin{equation}
        f_{D_s^*} = 0.274 \pm 0.006\,\GeV\,.
    \end{equation}
    The tensor decay constant is even less well known, and no lattice QCD results exist.
    We use a QCD sum rule determination of its ratio with respect to the pseudoscalar decay constant~\cite{Pullin:2021ebn}
    \begin{equation}
        \frac{f_{D_s^{*,T}}}{f_{D_s}}\bigg|_\text{QCDSR} = 1.13 \pm 0.07\,.
    \end{equation}
    For the reader's convenience, this translates to the approximate constraint
    \begin{equation}
        f_{D_s^{*,T}} \simeq 0.28 \pm 0.02\,\GeV\,.
    \end{equation}
    The total number of nuisance parameters is $2$.
    
    \item[$\boldsymbol{D\to \bar{K}\ell^+\nu}$] Assuming isospin symmetry, semileptonic $D^{+(0)} \to \bar{K}^{0(-)} \ell^+ \nu$ decays
    are described by three independent hadronic form factors $f_+^{D\to \bar{K}}(q^2)$,
    $f_0^{D\to \bar{K}}(q^2)$, and $f_T^{D\to \bar{K}}(q^2)$.
    We parametrise these form factors within the dispersively-bounded series expansion
    shown in \cref{eq:hme:ff_param},
    where each form factor is parametrised in a series of polynomials in $z(q^2)$, and we truncate the series at order $K=3$.
    Our choice of parametrisation ensures the identity $f_+^{D\to \bar{K}}(q^2 = 0) = f_0^{D\to \bar{K}}(q^2 = 0)$, thereby reducing the number of independent parameters by 1. \\
    
    Our analysis uses lattice QCD results by the HPQCD collaboration~\cite{Parrott:2022rgu} for all three form factors,
    which are provided as parameters of the BCL parametrisation~\cite{Bourrely:2008za}.
    Since our parametrisation \cref{eq:hme:ff_param} differs from the BCL one, we reconstruct the form factors and their correlations for all
    three form factors at three different values of the momentum transfer $q^2$.
    We remove one of these points for the form factor $f_+^{D\to \bar{K}}$ due to the exact relation between $f_+^{D\to \bar{K}}$ and $f_0^{D\to \bar{K}}$ at $q^2 = 0$.
    We further use lattice QCD results by the FNAL/MILC collaboration~\cite{FermilabLattice:2022gku}%
    \footnote{
    The HPQCD and FNAL/MILC analyses \cite{Parrott:2022rgu, FermilabLattice:2022gku} have a partial overlap in the ensembles used in their calculation.
    As the number of configurations used in the shared ensembles differs,
    we estimate that the correlation between these two sets of results can be safely neglected.
    }
    for the two form factors $f_+^{D\to \bar{K}}$ and $f_0^{D\to \bar{K}}$,
    which are also provided in terms of the BCL parameters but with a different truncation order than the HPQCD results.
    We also reconstruct the form factors and their correlations for both
    form factors at four different values of the momentum transfer $q^2$.
    Again, the exact form factor relation at $q^2 = 0$ allows us to remove one of these $q^2$ points.
    Besides the results of the FNAL/MILC and HPQCD collaborations,
    results from the ETM collaboration are also available~\cite{Lubicz:2017syv,Lubicz:2018rfs}. These results provide a total
    of 8 data points across all three form factors. We comment on the compatibility between the different lattice determinations below.

    The total number of nuisance parameters is $3(K + 1) - 1 = 11$.
    
    \item[$\boldsymbol{\Lambda_c\to \Lambda\ell^+\nu}$] Semileptonic $\Lambda_c^+\to\Lambda^0$ decays are described by ten independent hadronic form factors;
    six of these describe (axial)vector and (pseudo)scalar currents, and four further form factors describe tensor currents.
    Our analysis uses lattice QCD results from Ref.~\cite{Meinel:2016dqj} for the (axial)vector and (pseudo)scalar form factors. To our understanding,
    no lattice QCD results are presently available for the tensor form factors.
    Therefore, we use approximate relations between the tensor and the (axial)vector form factors for the same polarisation states. These relations arise from HQET and SCET symmetry considerations as
    discussed in \cref{app:hme:baryonic-ff-relations}.
    Similar to the $D\to \bar{K}$ form factors, each form factor is parametrised in a series expansion in $z(q^2)$ polynomials; see \cref{eq:hme:ff_param}.
    We choose to truncate the series at order $K=2$ since we find that the form factor uncertainties arising at this order are virtually
    indistinguishable from those obtained at $K=3$.
    Even for $K=2$, our lack of constraints on the tensor form factors implies that our parameter space is only bounded
    by the dispersive bounds.
    Four equations of motion and one algebraic identity lead to a total of five exact relations among the form factor parameters~\cite{Blake:2022vfl},
    reducing the number of independent parameters by 5.
    
    The total number of nuisance parameters is $10(K + 1) - 5 = 25$.
\end{description}

\begin{figure}[t!]
  \begin{minipage}[c]{0.48\textwidth}
    \includegraphics[width=\textwidth]{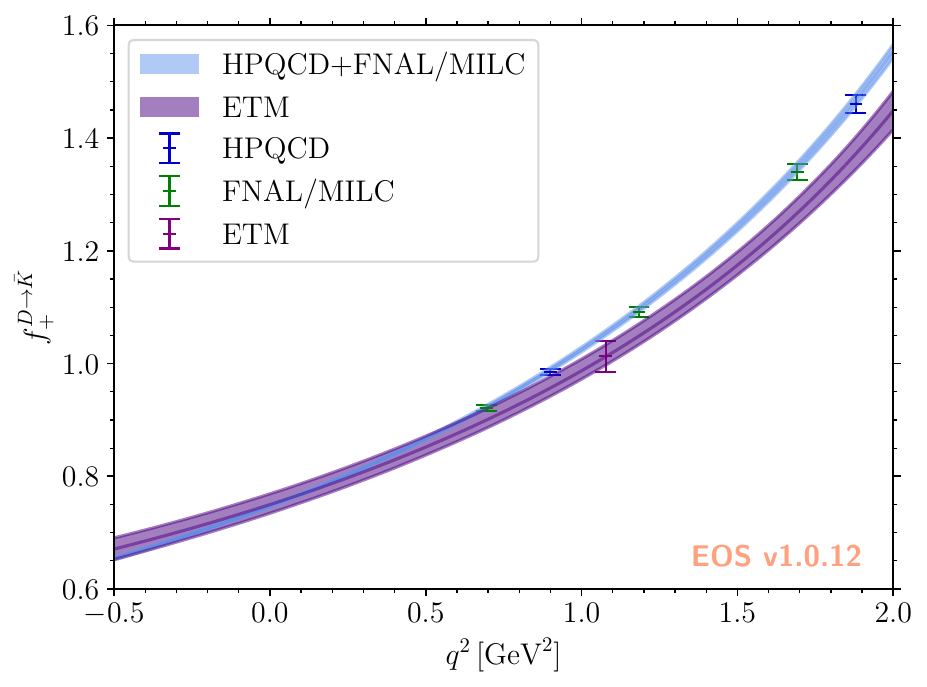}
  \end{minipage}\hfill
  \begin{minipage}[c]{0.48\textwidth}
    \includegraphics[width=\textwidth]{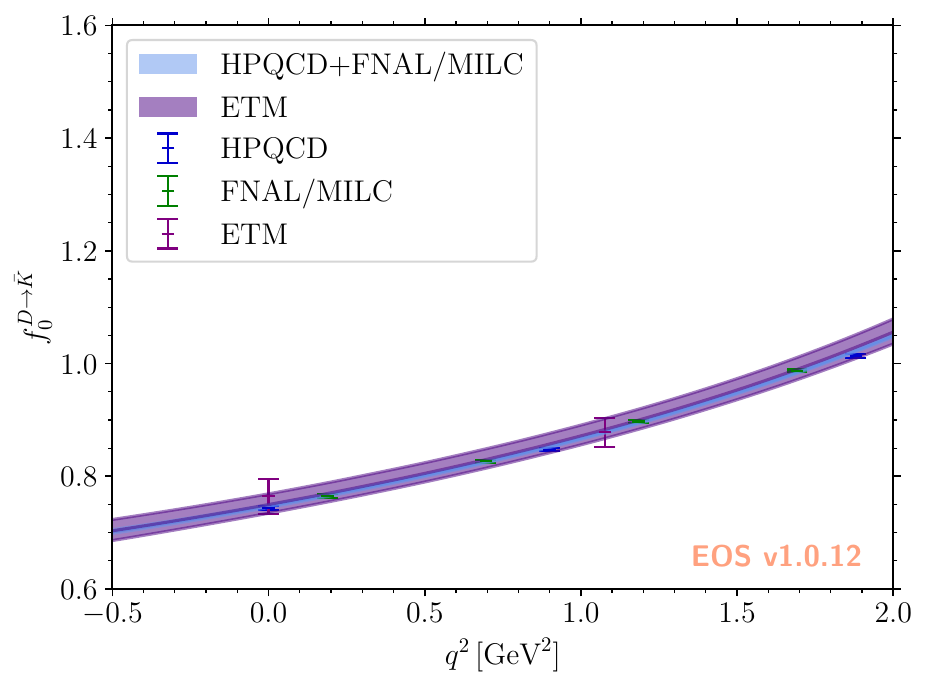}
  \end{minipage}
  
  \begin{minipage}[c]{0.48\textwidth}
    \includegraphics[width=\textwidth]{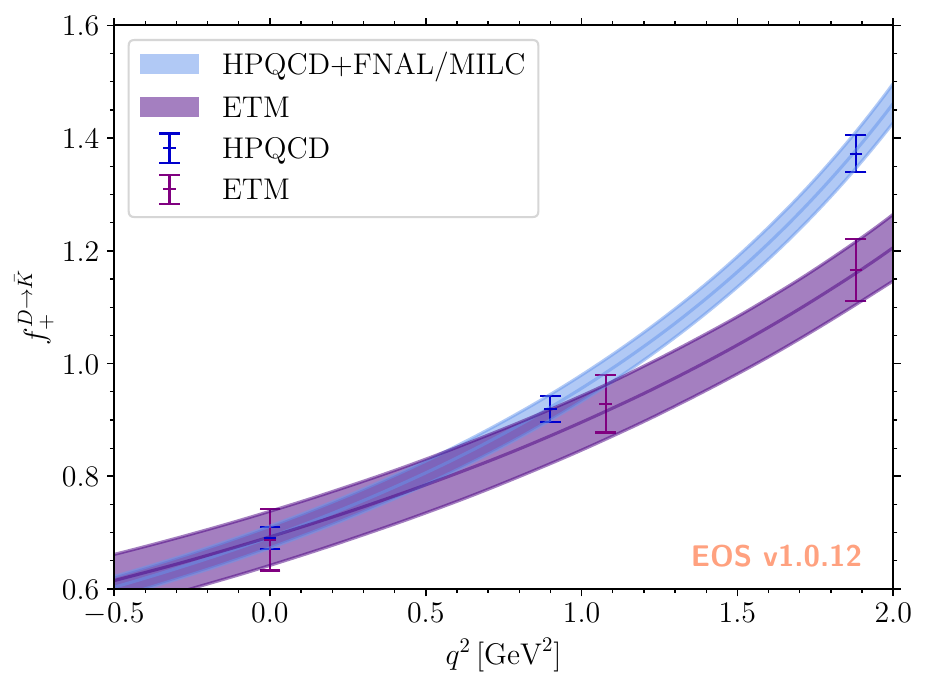}
  \end{minipage}\hfill
  \begin{minipage}[c]{0.48\textwidth}
    \caption{%
        Prior-predictive distributions for the $D\to \bar{K}$ form factors $f_+$, $f_0$, and
        $f_T$ as functions of the squared momentum transfer $q^2$ at truncation order $K=3$.
        The 68\% probability envelopes for the ``nominal'' results are shown as light-blue bands,
        the envelopes for the ETM results are shown as purple bands; see the discussion in
        \cref{sec:setup:nuisances}.
        The data points by the ETM~\cite{Lubicz:2017syv,Lubicz:2018rfs}, FNAL/MILC~\cite{FermilabLattice:2022gku}, and HPQCD~\cite{Parrott:2022rgu}
        collaborations are overlaid.
    }
    \label{fig:ff:dtok}
    \end{minipage}
\end{figure}

\paragraph{Comment on the $D\to \bar{K}$ form factor predictions}
\label{sec:comff}

Before proceeding with our analysis, we discuss the mutual compatibility of the individual $D\to \bar{K}$ lattice form factor results.
The 2021 FLAG average~\cite{FlavourLatticeAveragingGroupFLAG:2021npn} includes the ETM results~\cite{Lubicz:2017syv} and since-superseded HPQCD results~\cite{Chakraborty:2021qav} and
shows a small tension at the $2\,\sigma$ level between the two form factor determinations.
The average is dominated by the HPQCD results.
There is currently no FLAG average that includes the new determinations by FNAL/MILC~\cite{FermilabLattice:2022gku} and HPQCD 2022~\cite{Parrott:2022rgu}.

A simultaneous fit with our parametrisation to both HPQCD and FNAL/MILC constraints yields acceptable agreement,
with a total $\chi^2 = 9.8$ for $4$ degrees of freedom and a $p$ value of about $4\%$, above our a-priori threshold of $3\%$. 
However, the ETM results are not mutually compatible with the FNAL/MILC and HPQCD results, producing a
total $\chi^2 = 63.46$ for $10$ degrees of freedom in a simultaneous fit ($p$ value below $10^{-9}$).
We show the data points corresponding to all three lattice QCD results in \cref{fig:ff:dtok}.
The tension between ETM on the one hand and FNAL/MILC \& HPQCD on the other hand is clearly visible,
especially at $q^2$ close to its maximum value.%
\footnote{%
    \label{fn:experimental_shapes}
    We also note that the ETM-predicted differential rates agree very poorly with the experimental distributions.
    Comparison plots are provided in the supplementary material~\cite{EOS-DATA-2024-01}.
}

In case of the combination of univariate Gaussian distributions with substantial tensions, the PDG rescales the individual uncertainties with a scale factor
\begin{equation}
    \label{eq:scale_factor}
    S^2 \equiv \frac{\chi^2}{N_\text{d.o.f.}}\,.
\end{equation}
In principle, this recipe does not apply to our case since we aim to combine three \emph{multivariate} Gaussian distributions.
Nevertheless, we adopt the PDG procedure since there is no standard procedure for our case.
We obtain $S^2 = 6.346$.
Such a large value further indicates that the ETM and FNAL/MILC \& HPQCD results are incompatible and warrant further investigations on the lattice side.

This substantial tension between the individual lattice QCD results leads us to defining our ``nominal'' scenario,
in which we assume that the ETM results are an outlier and, therefore, drop them entirely.
Hence, we exclusively use the combination of FNAL/MILC and HPQCD results.

In addition, we consider a conservative scenario labelled ``scale factor'' defined to study the impact of removing the ETM results.
In this scenario, we assume that the ETM, FNAL/MILC, and HPQCD determinations all underestimate their respective uncertainties by a common factor.
We therefore adjust them by rescaling all three covariance matrices by the factor $S^2=6.346$.

\section{Methods and Results}
\label{sec:results}

\begin{table}[t]
\renewcommand{\arraystretch}{1.2}
\centering
\begin{tabular}{lc @{\hskip 3em} ccc @{\hskip 2em} c}
    \toprule
    Scenario & Fit model $M$ & $\chi^2$  & d.o.f.  & $p$ value [\%] & $\ln P(D, M)$ \\
    \midrule
    \multirow{3}{*}{nominal}
             & SM            & $60.9$    & $51$    & $16.1$         & $240.3 \pm 0.3$ \\
             & CKM           & $51.7$    & $50$    & $40.9$         & $251.7 \pm 0.3$ \\
             & WET           & $48.4$    & $42$    & $23.1$         & $250.3 \pm 0.3$ \\
    \midrule
    \multirow{3}{*}{scale factor}
             & SM            & $67.4$    & $51$    &  $6.2$         & $232.8 \pm 0.3$ \\
             & CKM           & $48.2$    & $50$    & $54.5$         & $249.1 \pm 0.3$ \\
             & WET           & $46.6$    & $42$    & $29.0$         & $248.7 \pm 0.3$ \\
   \bottomrule
\end{tabular}
\renewcommand{\arraystretch}{1.0}
\caption{%
    Goodness-of-fit values for the three main fits conducted as part of this analysis.
    We provide $\chi^2 = -2 \ln P(\text{experimental data}\,|\,\vecth^*, \vecnu^*)$ at the posterior's best-fit point
    $(\vecth^*, \vecnu^*)$ next to the $p$ value and the natural logarithm of the evidence $\ln P(D, M)$.
}
\label{tab:results:gof}
\end{table}

\begin{figure}[t]
    \centering
    \includegraphics[height=.6\textheight]{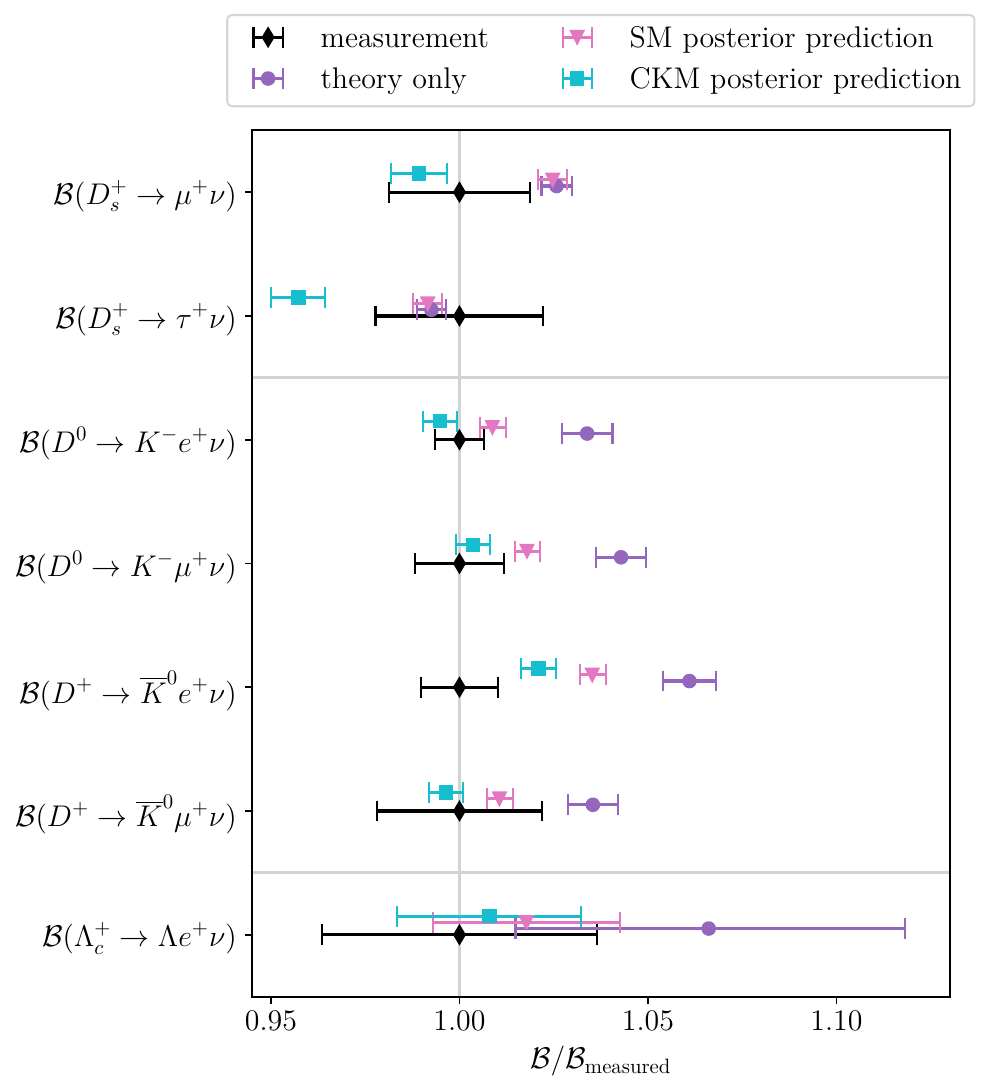}
    \caption{%
        Comparison of the predictions in our ``nominal'' scenario for the various branching ratios for $D_s^+\to \ell^+\nu$, $D^{+(0)}\to K^{0}_S(K^-)\ell^+\nu$,
        and $\Lambda_c\to \Lambda\ell^+\nu$.
        The decay $D_s^{*+}\to \ell^+\nu$ is omitted since the large theory uncertainties make a meaningful visual comparison
        off the predictions with the measurement impossible.
        Measurements are shown as black diamonds.
        Prior predictions labelled as ``theory only'' are shown as purple coloured circles.
        Posterior predictions for the two fit models are shown as magenta triangles (SM) and cyan squares (CKM).
    }
    \label{fig:results:comparison-BRs}
\end{figure}

We pursue three objectives with our analysis, which can be summarised
by the following questions:
\begin{enumerate}
    \item[(a)] Can the available data on exclusive $c\to s\ell^+\nu$ processes be described jointly
    by a single value for the CKM matrix element $\absVcs$, while simultaneously respecting the dispersive bounds?

    \item[(b)] Is a BSM/WET interpretation of the data favoured or disfavoured with respect to the SM hypothesis?

    \item[(c)] How strongly does the available data restrict the parameter space of the $sc\nu\ell$ sector of the WET?
\end{enumerate}
To achieve objective (a), we maximise the posterior PDFs $P(\vecth, \vecnu \,|\, D, M_i)$ with respect to
both the parameters of interest $\vecth$ and the nuisance parameters $\vecnu$. We do this for each of the
fit models $M_i \in \lbrace \text{SM}, \text{CKM}, \text{WET}\rbrace$ described in \cref{sec:setup:models}.
We compile the global $\chi^2$ values and their corresponding $p$ values for each of the best-fit points, $(\vecth^*, \vecnu^*)$, in \cref{tab:results:gof}.
\\
To achieve objective (b), we sample from the three posterior PDFs and calculate the marginal posteriors (or evidences)
$P(D \,|\, M_i)$. We compile the latter in \cref{tab:results:gof}. Calculating the marginal posteriors enables us
to carry out a Bayesian model comparison as discussed in \cref{sec:setup}.
\\
To achieve objective (c), we investigate the marginal posteriors for the WET parameters.
The marginal posteriors are discussed in detail in \cref{sec:results:BSM}.

The above steps are completed using \EOS~\cite{EOSAuthors:2021xpv}, a public software for flavour physics phenomenology.
This software provides numerical implementations for the theory predictions of observables arising in
leptonic $D_s^{*+} \to \ell^+\nu$ decays and semileptonic $D\to K\ell^+\nu$ and $\Lambda_c^+\to \Lambda^0(\to p \pi^-)\ell^+\nu$
decays. The predictions for $D_s^+\to \ell^+\nu$ and $D\to K\ell^+\nu$ are adapted from the expressions provided
in Ref.~\cite{Duraisamy:2014sna}. The predictions for $D_s^{*+} \to \ell^+\nu$ are adapted from the expressions
provided in Ref.~\cite{Plakias:2023esq} for $V\to \ell_1 \ell_2$. The predictions for
$\Lambda_c^+\to \Lambda^0(\to p \pi^-)\ell^+\nu$ are adapted from the expressions provided in Ref.~\cite{Boer:2019zmp}.
Predictions for all of these decays are possible with \EOS version 1.0.12 or newer~\cite{EOS:v1.0.12}.
To sample from the posterior density, \EOS uses dynamical nested sampling~\cite{Higson:2018}. To this end,
\EOS interfaces with the \dynesty software~\cite{Speagle:2020,dynesty:v2.0.3}.

\subsection{SM Prior Predictions and Fit}
\label{sec:results:SM}


Using the PDG reference value $\absVcs = 0.975$~\cite{PDGCKMReview},
we produce prior samples and prior-predictive distribution for the (pseudo)observables relevant to our analysis.
These prior samples, plots of the resulting hadronic form factors, and further plots are publicly available~\cite{EOS-DATA-2024-01}.

We find that our prior predictions for the integrated branching ratios systematically overshoot the measurements; the single exception is $\mathcal{B}(D_s^+\to \tau^+\nu)$.
This can be seen in \cref{fig:results:comparison-BRs}, where these predictions are labelled ``theory only'' and compared to the experimental data.\footnote{%
    For the production of \cref{fig:results:comparison-BRs}, we use the ``nominal'' fit scenario.
    We do not show the outcome of the ``scale factor'' scenario since the qualitative picture (overshooting the measurements)
    remains the same, although the tensions are somewhat reduced due to the inflated theory uncertainties.
}
To quantify this observation, we perform a $\chi^2$ test in our ``nominal'' fit scenario.
The agreement between our prior prediction and the measurements corresponds to a total $\chi^2 \simeq 133$
for $51$ degrees of freedom, corresponding to a tension of $5.9\,\sigma$; we refrain from producing a $p$ value.
This substantial tension is driven by the very precise measurements of the $D^{+(0)}\to \bar{K}^{0(-)} \ell^+\nu$
branching ratios, with individual tensions of $3.8\,\sigma$ for $D^0\to K^- \mu^+\nu_\mu$, $5.4\,\sigma$ for $D^0\to K^- e^+\nu_e$ and $6\,\sigma$ for $D^+\to \bar{K}^0 e^+\nu_e$ and one degree of freedom each.
In contrast, the agreement between the prior predictions and measurements for the kinematic distribution
$d\Gamma/dq^2$ in these semileptonic decays is very good, with $\chi^2 / \text{d.o.f.} = 4.5/8$
for $D^+\to \bar{K}^0e^+\nu$, $21.4/17$ for $D^0\to \bar{K}^-e^+\nu$, and $17.9/17$ for $D^0\to K^-\mu^+\nu$.

The situation improves slightly in the ``scale factor'' model.
The global $\chi^2$ reduces to $108$ for the same $51$ degrees of freedom, \ie a $4.6\,\sigma$ tension.
However, the reduction in the tension of the individual branching ratios (now $2.8\,\sigma, 3.6\,\sigma$ and $4.9\,\sigma$ respectively)
is compensated by larger tensions in the kinematic distributions, as anticipated by the discussion in \cref{fn:experimental_shapes}.\\

These observations suggest a possible problem in the normalisation of $D\to \bar{K}\ell^+\nu$ decays.
This problem could stem either from issues in the normalisation of the lattice QCD results for the $D\to \bar{K}$ form factors (discussed below);
from using an incorrect value of \absVcs (discussed in \cref{sec:results:CKM});
or from experimental issues in measuring the absolute branching fractions (beyond the scope of this work).

To test a possible issue with the normalisation of the lattice QCD results,
we float all hadronic nuisance parameters within the SM fit model (``nominal'' scenario) and fit to the full experimental likelihood.
We find that we can reduce the experimental $\chi^2$ to $60.9$ with a $p$ value of $16.1\%$.
This happens at the expense of moving away from the a-priori parameter values by $4.0\,\sigma$.
We produce posterior-predictive distributions for the relevant absolute branching fractions, which are shown
in \cref{fig:results:comparison-BRs} as magenta triangles.
We can still observe that the posterior predictions systematically overshoot the measurements.
However, the tensions with respect to the experimental measurements are visibly reduced.
At the level of the variety, number, and accuracy of the hadronic matrix elements used in this analysis,
we find compatibility between the hadronic matrix elements and the dispersive bounds.
\\

The same qualitative behaviour is observed within the ``scale factor'' scenario, albeit with reduced tensions due to the inflated theory uncertainties.

Our findings strongly support the notion that the extraction of $\absVcs$ should only be undertaken within a global fit of the available data.

\begin{figure}[t]
    \centering
    \includegraphics[width=.49\textwidth]{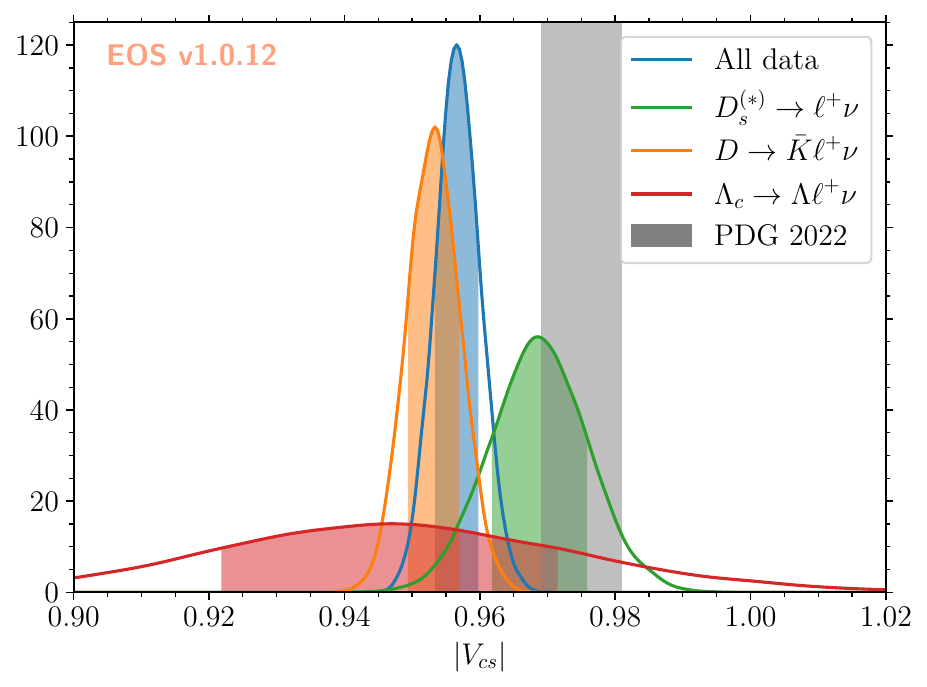} \hfill
    \includegraphics[width=.49\textwidth]{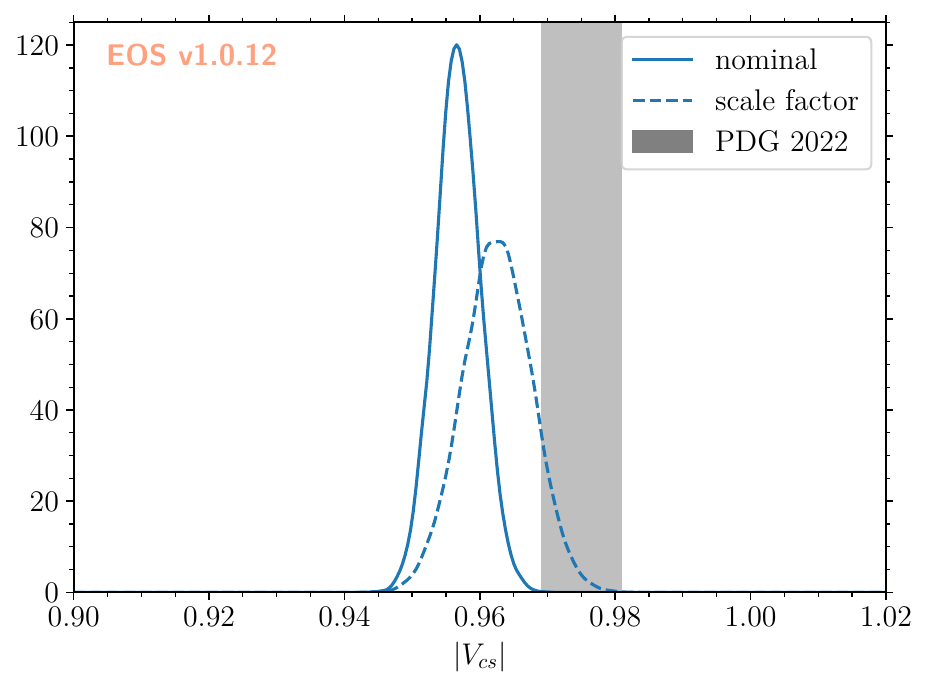} \hfill
    \caption{%
        Marginalised one-dimensional posterior densities for $|V_{cs}|$ within the CKM fit model.
        We show our nominal result for the full data set as described in \autoref{sec:setup:exp} in blue.
        Additional results for datasets only containing either
        $D_s^{(*)+}\to \ell^+\nu$, $D^{0(+)}\to \bar{K}^{-(0)}\ell^+\nu$, or $\Lambda_c\to \Lambda \ell^+\nu$
        data are shown in the figure on the left in green, orange and red, respectively.
        The shaded areas indicate the central intervals at $68\%$ probability.
        The figure on the right presents the result for the full data set in the ``nominal'' and ``scale factor'' scenarios. 
    }
    \label{fig:results:CKM:marg-Vcs}
\end{figure}

\subsection{CKM Fit}
\label{sec:results:CKM}
We now determine $\absVcs$ from individual fits to $D_{s}^{(*)+}\to \ell^+\nu$, $D\to \bar{K}\ell^+\nu$, $\Lambda_c\to\Lambda\ell^+\nu$
and the combination of all these decays.
The resulting distributions of the extracted values for $|V_{cs}|$ are shown on the left-hand side of \cref{fig:results:CKM:marg-Vcs}.
A summary of the obtained \absVcs values and the individual goodness-of-fit diagnostics are shown in \cref{tab:results:CKM}.
\\

\begin{table}[t]
    \renewcommand{\arraystretch}{1.2}
    \centering
    \begin{tabular}{lc @{\hskip 0em} D{.}{.}{7.2} D{.}{.}{2.0} D{.}{.}{5.1} @{\hskip 0em} c}
        \toprule
        Scenario & Data set             & \multicolumn{1}{c}{$\chi^2$}
                                                  & \multicolumn{1}{c}{d.o.f.}
                                                            & \multicolumn{1}{c}{$p$ value [\%]}
                                                                             & \multicolumn{1}{c}{$|V_{cs}|$}
        \\
        \midrule
        & $D_s^{(*)+}\to \ell^+\nu$     &    $2.5$  &  $2$    & $28.0$         & $0.969 \pm 0.007$
        \\
        & $\Lambda_c\to \Lambda\ell\nu$ &    $0.1$  &  $1$    & $81.2$         & $0.947 ^{+0.027}_{-0.026}$
        \\
        \midrule
        \multirow{2}{*}{nominal}
        & $D\to \bar{K}\ell\nu$
                                        &   $44.1$  & $45$    & $50.9$         & $0.953 \pm 0.004$
        \\
        & joint fit                     &   $51.7$  & $50$    & $40.9$         & $0.957 \pm 0.003$
        \\
        \midrule
        \multirow{2}{*}{scale factor}
        & $D\to \bar{K}\ell\nu$         &   $42.7$  & $45$    & $57.0$         & $0.957 \pm 0.007$
        \\
        & joint fit                     &   $48.2$  & $50$    & $54.5$         & $0.963 \pm 0.005$
        \\
        \bottomrule
    \end{tabular}
    \renewcommand{\arraystretch}{1.0}
    \caption{%
        Goodness-of-fit values and results for the individual CKM fits discussed in \cref{sec:results:CKM}.
        The results for $|V_{cs}|$ represent the median values and central $68\%$ probability intervals of the marginal 1D
        posterior probability densities, which we find to be symmetric.
    }
    \label{tab:results:CKM}
\end{table}

The result of our joint fit in the ``nominal'' scenario is
\begin{equation}
\label{eq:results:Vcs}
    |V_{cs}| = 0.957 \pm 0.003\,,
\end{equation}
which agrees with the weighted average of the individual fit results.
The central value for the ``scale factor'' result is higher than the above but compatible at the $1\,\sigma$ level, as shown on the right-hand side of  \cref{fig:results:CKM:marg-Vcs}.
As illustrated in \cref{fig:results:comparison-BRs}, floating the CKM parameter $|V_{cs}|$
results in overall better agreement with the data.
Contrary to the SM prior predictions and the SM fit posterior predictions,
no systematic shift to larger or smaller values for the branching ratios is visible.
This is also reflected in \cref{tab:results:gof}.
For the ``nominal'' scenario, we show that our CKM fit yields a reduction of the $\chi^2$ in the best-fit point
by $9$ at the expense of one degree of freedom. Using Wilks' theorem, we therefore obtain a preference for the CKM fit
over the SM fit at the $2.7\,\sigma$ level. A Bayesian model comparison yields a Bayes factor of $\simeq 8.9\cdot 10^4$,
decisively in favour of the CKM fit model.
For the ``scale factor'' scenario, we find a similar preference for the CKM fit model.

We find that the dispersive bounds only mildly affect the extraction of \absVcs.
Their main effect is a reduction of the uncertainties of the $\Lambda_c \to \Lambda$ form factors,
which play only a secondary role in the global fit due to the large experimental and theory uncertainties.
The posterior-predictive distributions of the saturations in all our models are available as supplementary material~\cite{EOS-DATA-2024-01}.\\

Our nominal result is compatible with the PDG reference value~\cite{PDGCKMReview}
\begin{equation}
\label{eq:results:CKM:world-avg}
    |V_{cs}|_\text{PDG}
        = 0.975 \pm 0.006\,,
\end{equation}
at the level of $2.7\,\sigma$ (nominal) and $1.5\,\sigma$ (scale factor). The shift away from the PDG reference value can be understood as follows:
\begin{itemize}
    \item We account
    for universal electroweak corrections as part of the Wilson coefficients in our theory predictions. The correction is commonly known
    as the Sirlin factor~\cite{Sirlin:1980nh} and defined in \eqref{eq:setup:WET:WCsir}.\\
    The PDG result~\cite{PDGCKMReview}
    extracted from both $\mathcal{B}(D_s^+\to \mu^+\nu)$ and $\mathcal{B}(D_s^+\to \tau^+\nu)$ reads
    \begin{equation}
        \absVcs_{\text{PDG,$D_s\to \ell^+\nu$}} = 0.984 \pm 0.012\,.
    \end{equation}
    However, we can only reproduce this partial result
    if we do not account for the Sirlin factor. Including this factor would lower the PDG partial result to $0.974$,
    much closer to our result for these leptonic modes quoted in \cref{tab:results:CKM}.

    \item The PDG result~\cite{PDGCKMReview} 
    for $D\to \bar{K}\ell^+\nu$ 
    \begin{equation}
        \absVcs_{\text{PDG,$D\to\bar{K}\ell^+\nu$}} = 0.972 \pm 0.007
    \end{equation}
    is based on the hadronic form factor evaluated at $q^2 = 0$, $f_+(0) = 0.7385 \pm 0.0044$.
    This form factor value is obtained by FLAG~\cite{FlavourLatticeAveragingGroupFLAG:2021npn}
    from ETM~\cite{Lubicz:2017syv} and since superseded HPQCD~\cite{Chakraborty:2021qav} results.
    We obtain $f_+(0) = 0.747\pm 0.002$ in our nominal model and $f_+(0) = 0.744 \pm 0.005$ in the scale factor scenario.
    Both values are larger than the FLAG value by about $1\%$.
    Adjusting for our results and applying the Sirlin factor, the PDG value would shift downward by about $2\%$
    to $\absVcs = 0.952$, bringing the PDG value in good alignment with our partial results in \cref{tab:results:CKM}.
\end{itemize}
We therefore conclude that a substantial fraction of the observed shift in $\absVcs$ is due to our
inclusion of the Sirlin factor.

\subsection{CKM unitarity}
\label{sec:results:CKM-unitarity}

The violation of CKM unitarity of around $3\,\sigma$ by the first-row elements of the CKM matrix has received quite some attention \cite{Seng:2018yzq,Belfatto:2019swo,Grossman:2019bzp,Crivellin:2020lzu,Kirk:2020wdk,Crivellin:2021njn,Seng:2021nar,Cirigliano:2022yyo,Crivellin:2022rhw,Cirigliano:2023nol}.
Here, we perform an alternative test of
CKM unitarity by probing the normalisation of both the second row and column;
\begin{align}
    2^{\rm nd}{\rm \; row:} \; \sum_{D = d,s,b} |V_{cD}|^2\, \quad\quad\quad 2^{\rm nd}{\rm \; column:} \;\sum_{U = u,c,t} |V_{Us}|^2 
\end{align}

We test this using the PDG reference value for $\absVcs$ in \eqref{eq:results:CKM:world-avg},  our determination in \eqref{eq:results:Vcs} and the ``scale factor'' determination. For the other CKM elements, we use the present PDG reference values
~\cite{PDGCKMReview}
\begin{equation}
\begin{aligned}
    |V_{cd}|^\text{PDG}
        & = 0.221 \pm 0.004\,, &
    |V_{cb}|^\text{PDG}
        & = (40.8 \pm 1.4) \times 10^{-3}\,,&
\end{aligned}
\end{equation}
and
\begin{equation}
\begin{aligned}
    |V_{us}|^\text{PDG}
        & = 0.2243 \pm 0.0008\,, &
    |V_{ts}|^\text{PDG}
        & = (41.5 \pm 0.9) \times 10^{-3}\,.&
\end{aligned}
\end{equation}
Our results are given in \cref{tab:CKM_unitarity}. Under the assumption of $100\%$ positively correlated uncertainties,
we find a deficit toward the expectation of CKM second-row and second-column unitarity at the $4.3\,\sigma$ level and $5.2\,\sigma$ level using the nominal results.

\begin{table}[t]
    \centering
    \begin{tabular}{l ccc}
        \toprule
                                & PDG                               & nominal                           & scale factor \\
         \midrule
         $\absVcs$              & $0.975 \pm 0.006$                 & $0.957 \pm 0.003$                 & $0.963 \pm 0.005$ \\
         $2^\text{nd}$ row      & $1.00 \pm 0.014 ~ (0.08\,\sigma)$ & $0.966 \pm 0.008 ~ (4.3\,\sigma)$ & $0.978 \pm 0.012 ~ (1.9\,\sigma)$ \\
         $2^\text{nd}$ column   & $1.00 \pm 0.012 ~ (0.22\,\sigma)$ & $0.968 \pm 0.006 ~ (5.2\,\sigma)$ & $0.979 \pm 0.010 ~ (2.0\,\sigma)$ \\
         \bottomrule
    \end{tabular}
    \caption{Results of the CKM unitarity tests.
    The last two lines contain the squared sum of the CKM elements of the second row or column, as well as the pull to unity assuming Gaussian uncertainties.
    The uncertainties on the CKM entries are assumed to be 100\% positively correlated, which corresponds to the most conservative scenario (proper estimations would give larger pulls).}
    \label{tab:CKM_unitarity}
\end{table}
This large tension with unitarity again strengthens the case to investigate the normalisation issue that we already pointed out in
\cref{sec:results:SM}.

\subsection{BSM Interpretation}
\label{sec:results:BSM}

\begin{figure}
    \hspace{-.1\textwidth}
    \includegraphics[width=1.2\textwidth]{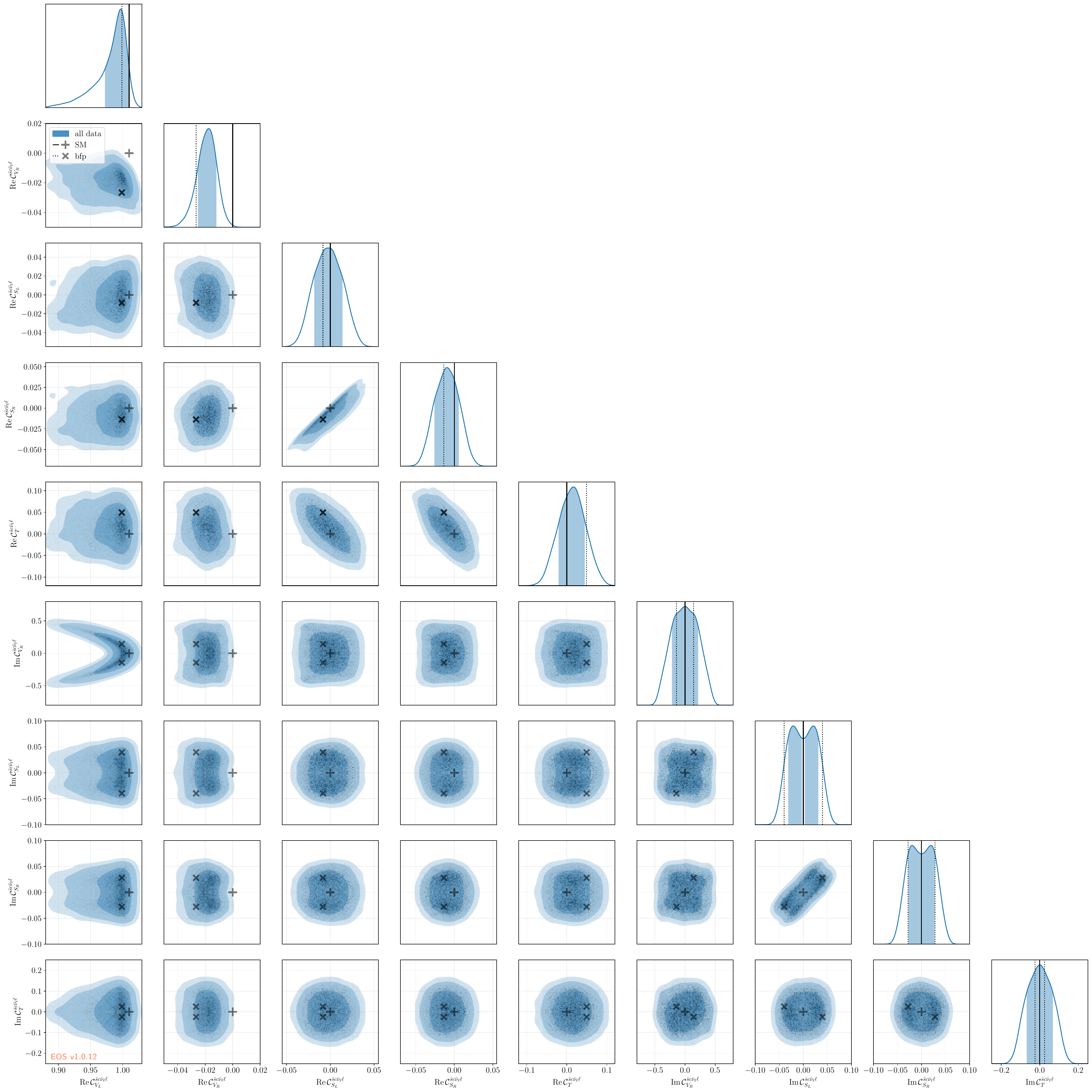}
    \caption{
        Marginalised 1D- and 2D-posterior distributions for the 9 parameters of interest of our WET model in the nominal scenario.
        As discussed in the text, the imaginary part of \wc[]{V_L} is set to zero using the global unconstrained phase.
        The ``+'' and the solid black lines show the SM point $\wc[]{V,L} = 1.01$ and $\wc[]{i} = 0$ for all other operators.
        The ``$\times$'' and the dashed black lines show the position of our best-fit point.
        The blue areas are the central 68\%, 95\%, and 99\% integrated probability contours of the posterior distribution obtained from a kernel density estimation.
    }
    \label{fig:results:WET-corner-plot}
\end{figure}

Our findings so far motivate us to investigate further the allowed parameter space for BSM contribution to $c\to s\ell^+\nu$ processes. Lifting the assumption
of SM dynamics, we fit the 9 parameters discussed in \cref{sec:setup:models}.
We find the resulting posterior PDF to be multi-modal and each mode to be distinctly
non-Gaussian. The individual modes are related through symmetries of our likelihood
and therefore feature the same maximum a-posteriori.
Here and in our supplementary material, we provide information on one chosen mode
of the posterior, which is defined by $\arg \wc{V,L} = 0$ and $\wc{V,L} \simeq 1$.
In our nominal scenario, we obtain the following 68\% probability intervals 
\begin{equation}
\begin{aligned}
    \Re \wc{V,L}
        & = [\phantom{-}0.957, \phantom{-}1.002]\,,
    &
    \\
    \Re \wc{V,R}
        & = [-0.026, -0.012]\,,
    &
    \Im \wc{V,R}
        & = [-0.225, 0.225]\,,
    \\
    \Re \wc{S,L}
        & = [-0.019, \phantom{-}0.014]\,,
    &
    \Im \wc{S,L}
        & = [-0.030, 0.030]\,,
    \\
    \Re \wc{S,R}
        & = [-0.026, \phantom{-}0.006]\,,
    &
    \Im \wc{S,R}
        & = [-0.028, 0.028]\,,
    \\
    \Re \wc{T}
        & = [-0.021, \phantom{-}0.046]\,,
    &
    \Im \wc{T}
        & = [-0.068, 0.068]\,.
\end{aligned}
\end{equation}
This mode and its 68, 95 and 99\% central probability intervals are shown in \autoref{fig:results:WET-corner-plot}, together with the SM and best-fit points.
The ``scale factor'' scenario yields qualitatively the same results.
While our results indicate very strong constraints on potential (pseudo)scalar and tensor effects in $sc\nu\ell$ sector of the Weak Effective Theory,
they do allow for surprisingly large CP-violating effects in right-handed currents,
at the level of $23\%$ of the SM value for the left-handed current.
Such a new source of CP violation in the second quark generation would, of course, be interesting in the context of the observed CP asymmetry in
non-leptonic $D$ decays~\cite{LHCb:2019hro}.
However, recent analyses of the high $p_T$ lepton tails in Drell-Yan processes~\cite{Fuentes-Martin:2020lea} seem to exclude this type of explanation.

\section{Conclusion}
\label{sec:conclusion}

We have performed a comprehensive global analysis of $c\to s\ell^+\nu$ decays within the Standard Model (SM) of particle physics
and within the Weak Effective Theory (WET).
Our analysis is the first to account for dispersive bounds that connect many of the hadronic parameters
needed in the description of the various decays.
We study the impact of the tension between the ETM and FNAL+MILC \& HPQCD lattice QCD results for the $D\to K$ form factors.
We found that enlarging the theory uncertainties of these quantities does not change our results qualitatively.
Moreover, our analysis includes for the first time data on leptonic $D_s^{*+}$ and semileptonic $\Lambda_c$ decays.\\

Assuming SM dynamics, our nominal fit yields
\begin{equation}
    |V_{cs}| = 0.957 \pm 0.003\,.
\end{equation}
Our result deviates from the reference values by the Particle Data Group by more than $2.7\,\sigma$.
A large part of the observed discrepancy is traced back to a different treatment of the electroweak corrections.
Our findings lead to a $4.3\sigma$ and $5.2\sigma$ deviation from unitarity in the second row and second column of the CKM quark-mixing matrix. 
In light of the observed tension between the lattice QCD results for the $D\to \bar{K}$ form factors, our results should be revised once this tension is clarified.\\

We set stringent constraints on potential BSM effects in the $sc\nu\ell$ sector of the WET, which limit
hypothetical (pseudo)scalar or tensor effects to be below $7\%$ of $\wc[\mathrm{SM}]{V,L} \simeq 1.01$,
the SM value for the Wilson coefficient of the left-handed operator.
Nevertheless, we find that right-handed currents can still be sizeable. In particular, we find that
CP-violating effects in right-handed currents at the level of $\simeq 23\%$ of the SM contribution
are not yet excluded.
Complementary experimental data, such as the angular distribution of $\Lambda_c\to\Lambda(\to p \pi)\ell^+\nu$
decays would provide the statistical power to exclude such large CP-violating right-handed contributions.

\acknowledgments

We thank Augusto Ceccucci, Zoltan Ligeti, and Yoshihide Sakai for useful communication
on the section ``CKM Quark-Mixing Matrix'' within the Particle Data Group's Review of
Particle Physics~\cite{PDGCKMReview}.
We would like to thank Admir Greljo, Matthew Kirk, and Wolfgang Altmannshofer for useful comments on the manuscript.
DvD~acknowledges support by the UK Science and Technology Facilities Council
(grant numbers ST/V003941/1 and ST/X003167/1). KKV acknowledges
support from the Dutch Research Council (NWO) in the form of the VIDI grant “Solving
Beautiful Puzzles”.

\appendix

\section{Treatment of the Hadronic Matrix Elements}
\label{app:hme}

To access quark-level properties of leptonic or semileptonic decays as we do in this analysis,
knowledge of the relevant hadronic matrix elements is essential. The latter parametrise the
mismatch between quark-level processes such as $c\to s\ell^+\nu$ and the hadronic processes
such as $D_s^{(*)}\to \ell^+\nu$, $D^{+(0)}\to \bar{K}^{0(-)}\ell^+\nu$, and $\Lambda_c^+\to \Lambda^0\ell^+\nu$.
As such, the hadronic matrix elements are genuinely nonperturbative objects that need to be
inferred, ideally, from first-principle methods such as lattice QCD. Where lattice QCD results
are unavailable, we fall back to QCD sum rule estimates.\\
For convenience, the hadronic matrix elements are typically expressed in terms of scalar-valued
hadronic decay constants or hadronic form factors. The latter distinguish themselves from the
decay constant by virtue of being scalar-valued functions of the momentum transfer. Throughout
this work, we denote the squared momentum transfer $q^2 = m_{\ell\nu}^2$.

\subsection{Definitions}
\label{app:hme:defs}

The simplest hadronic matrix element arises in the decay of a pseudoscalar $D_s^+$ state to a lepton-neutrino pair.
We use a common, if not the standard, definition of the decay constant~\cite{FlavourLatticeAveragingGroupFLAG:2021npn}.
\begin{equation}
\label{eq:hme:defs:psd-decay-const}
\begin{aligned}
    \bra{0} \bar{s} \gamma^\mu \gamma_5 c \ket{D_s^+(p)}
        & = i f_{D_s} p^\mu\,,
    &
    \bra{0} \bar{s} \gamma_5 c \ket{D_s^+(p)}
        & = -i \frac{M_{D_s}^2}{m_c(\mu_c) + m_s(\mu_c)} f_{D_s}\,,
\end{aligned}
\end{equation}
where the axial decay constant also describes the scale-dependent hadronic matrix element
of the pseudoscalar current.\\

The next-to-simplest case arises in the leptonic decay of a vector $D_s^{*+}$ meson.
We follow the convention of Ref.~\cite{RBC-UKQCD:2008mhs,Plakias:2023esq} and use the definition
\begin{equation}
\label{eq:hme:defs:vec-decay-consts}
\begin{aligned}
    \bra{0} \bar{s} \gamma^\mu c \ket{D_s^{*+}(p, \varepsilon)}
        & = f_{D_s^*} M_{D_s} \varepsilon^\mu\,,
    &
    \bra{0} \bar{s} \sigma^{\mu\nu} c \ket{D_s^{*+}(p, \varepsilon)}
        & = i f_{D_s^*}^T \left( \varepsilon^\mu p^\nu - p^\mu \varepsilon^\nu\right)\,.
\end{aligned}
\end{equation}
Note that in this case, the two non-vanishing matrix elements are not related by equations of motion
and, therefore, do not share a common decay constant.\\

The simplest set of three form factors arises in $D\to K\ell^+\nu$ decays from vector, scalar, and tensor currents.
A common definition of the form factors reads
\begin{align}
    \bra{K(k)} \bar{s} \gamma^\mu c \ket{D(p)}
        & = f^{D\to K}_+(q^2) \left[(p + k)^\mu - q^\mu \frac{M_D^2 - M_K^2}{q^2}\right]
          + f^{D\to K}_0(q^2) q^\mu \frac{M_D^2 - M_K^2}{q^2}\,,
    \\
    \bra{K(k)} \bar{s} c \ket{D(p)}
        & = f^{D\to K}_0(q^2) \frac{M_D^2 - M_K^2}{m_c(\mu_c) - m_s(\mu_c)}\,,
    \\
    \bra{K(k)} \bar{s} \sigma^{\mu\nu} q_\nu c \ket{D(p)}
        & = \frac{i f^{D\to K}_T(q^2)}{M_D + M_K} \left[ q^2 (p + k)^\mu - (M_D^2 - M_K^2) q^\mu \right] \,.
\end{align}
In the above, $q \equiv p - k$. The dependence of the form factors as functions of $q^2$ is discussed
in \cref{app:hme:bounds}.

The most complicated case, in our analysis, arises in $\Lambda_c\to \Lambda \ell^+\nu$ decays.
A common definition of the ten form factors reads, following the notation of Ref.~\cite{Boer:2014kda}, 
\begin{align}
    \bra{ \Lambda } \overline{s} \,\gamma^\mu\, c \ket{ \Lambda_c }
        & = \overline{u}_\Lambda(k,s_{\Lambda}) \bigg[ f_{V,t}^{\Lambda_c\to\Lambda}(q^2)\: (m_{\Lambda_c}-m_\Lambda)\frac{q^\mu}{q^2} \\
    \nonumber
        &   \phantom{\overline{u}_\Lambda \bigg[}+ f_{V,0}^{\Lambda_c\to\Lambda}(q^2) \frac{m_{\Lambda_c}+m_\Lambda}{s_+}
            \left( p^\mu + k^{ \mu} - (m_{\Lambda_c}^2-m_\Lambda^2)\frac{q^\mu}{q^2}  \right) \\
    \nonumber
        &   \phantom{\overline{u}_\Lambda \bigg[}+ f_{V,\perp}^{\Lambda_c\to\Lambda}(q^2)
            \left(\gamma^\mu - \frac{2m_\Lambda}{s_+} p^\mu - \frac{2 m_{\Lambda_c}}{s_+} k^{ \mu} \right) \bigg] u_{\Lambda_c}(p,s_{\Lambda_c}) \, , \\
    \bra{ \Lambda } \overline{s} \,\gamma^\mu\gamma_5\, c \ket{ \Lambda_c }
        & = -\overline{u}_\Lambda(k,s_{\Lambda}) \:\gamma_5 \bigg[ f_{A,t}^{\Lambda_c\to\Lambda}(q^2)\: (m_{\Lambda_c}+m_\Lambda)\frac{q^\mu}{q^2} \\
    \nonumber
        &   \phantom{\overline{u}_\Lambda \bigg[}+ f_{A,0}^{\Lambda_c\to\Lambda}(q^2)\frac{m_{\Lambda_c}-m_\Lambda}{s_-}
            \left( p^\mu + k^{ \mu} - (m_{\Lambda_c}^2-m_\Lambda^2)\frac{q^\mu}{q^2}  \right) \\
    \nonumber 
        & \phantom{\overline{u}_\Lambda \bigg[}+ f_{A,\perp}^{\Lambda_c\to\Lambda}(q^2) \left(\gamma^\mu + \frac{2m_\Lambda}{s_-} p^\mu - \frac{2 m_{\Lambda_c}}{s_-} k^{ \mu} \right) \bigg]  u_{\Lambda_c}(p_{\Lambda_c},s_{\Lambda_c}), \\
    \bra{ \Lambda } \overline{s} \,i\sigma^{\mu\nu} q_\nu \, b \ket{ \Lambda_c } 
        &= - \overline{u}_\Lambda(k,s_{\Lambda}) \bigg[  f_{T,0}^{\Lambda_c\to\Lambda}(q^2) \frac{q^2}{s_+} \left( p^\mu + k^{\mu} - (m_{\Lambda_c}^2-m_{\Lambda}^2)\frac{q^\mu}{q^2} \right) \\
    \nonumber 
        & \phantom{\overline{u}_\Lambda \bigg[} + f_{T,\perp}^{\Lambda_c\to\Lambda}(q^2)\, (m_{\Lambda_c}+m_\Lambda) \left( \gamma^\mu -  \frac{2  m_\Lambda}{s_+} \, p^\mu - \frac{2m_{\Lambda_c}}{s_+} \, k^{ \mu}   \right) \bigg] u_{\Lambda_c}(p,s_{\Lambda_c}) \, , \\
    \bra{ \Lambda } \overline{s} \, i\sigma^{\mu\nu}q_\nu \gamma_5  \, c \ket{ \Lambda_c }
        & = -\overline{u}_{\Lambda}(k,s_{\Lambda}) \, \gamma_5 \bigg[   
        f_{T5,0}^{\Lambda_c\to\Lambda}(q^2) \, \frac{q^2}{s_-}
            \left( p^\mu + k^{\mu} -  (m_{\Lambda_c}^2-m_{\Lambda}^2) \frac{q^\mu}{q^2} \right) \\
    \nonumber
        &   \phantom{\overline{u}_\Lambda \bigg[}  + f_{T5,\perp}^{\Lambda_c\to\Lambda}(q^2)\,  (m_{\Lambda_c}-m_\Lambda)
            \left( \gamma^\mu +  \frac{2 m_\Lambda}{s_-} \, p^\mu - \frac{2 m_{\Lambda_c}}{s_-} \, k^{ \mu}  \right) \bigg]  u_{\Lambda_c}(p,s_{\Lambda_c})\,,
\end{align}
where $s_{\Lambda_{(c)}}$ denotes the spin of the $\Lambda_{(c)}$, $p$ is the momentum of the $\Lambda_c$, $k$ is the momentum of the $\Lambda$,
and we abbreviate $s_{\pm} = (m_{\Lambda_b} \pm m_\Lambda)^2 - q^2$.

\subsection{Dispersive Bounds}
\label{app:hme:bounds}

\begin{table}
  \centering
  \renewcommand{\arraystretch}{1.4}
  \setlength{\tabcolsep}{10pt}
  \newcommand{\schii}[2]{\chi^{(#1)}_{#2}}
  \newcommand{\schi}[3]{\chi^{(#1)}_{#2}\big|_{#3}}
  \begin{tabular}{@{}l c c c c@{}}
    \toprule
    $\schii{\lambda}{\Gamma}$
                               & $n$ & $\schi{\lambda}{\Gamma}{\mathrm{OPE}}\times 10^2$
                                                      & Form factor                                           & Pole $R_F$ (mass, decay constant) $[\mathrm{GeV}]$ \\
    \midrule
    $\schii{J=0}{V}$             & $1$ & $1.38$         & $f_0^{D\to \bar{K}}$, $f_{t,V}^{\Lambda_c\to\Lambda}$ & --- \\
    $\schii{J=1}{V}$             & $2$ & $1.52/m_c^2$   & $f_+^{D\to \bar{K}}$, $f_{0,V}^{\Lambda_c\to\Lambda}$, $f_{\perp,V}^{\Lambda_c\to\Lambda}$ & $\bar{D}_s^* \, (2.112, 0.274(6))$ \\
    $\schii{J=0}{A}$             & $1$ & $2.51$         & $f_{t,A}^{\Lambda_c\to\Lambda}$                       & $\bar{D}_s   \, (1.968, 0.2499(5))$ \\
    $\schii{J=1}{A}$             & $2$ & $0.98/m_c^2$   & $f_{0,A}^{\Lambda_c\to\Lambda}$, $f_{\perp,A}^{\Lambda_c\to\Lambda}$                       & --- \\
    $\schii{J=1}{T}$             & $3$ & $1.12/m_c^2$   & $f_T^{D\to K}$, $f_{0,T}^{\Lambda_c\to\Lambda}$, $f_{\perp,T}^{\Lambda_c\to\Lambda}$       & $\bar{D}_s^* \, (2.112, 0.28(2))$ \\
    $\schii{J=1}{AT}$            & $3$ & $0.88/m_c^2$   & $f_{0,T5}^{\Lambda_c\to\Lambda}$, $f_{\perp,T5}^{\Lambda_c\to\Lambda}$                      & --- \\
    \bottomrule
  \end{tabular}
  \caption{
        \label{tab:hme:chi}
        List of the minimally-subtracted correlators $\schii{\lambda}{\Gamma}$ relevant to the hadronic matrix elements used in this analysis.
        We calculate the numerical values based on formulas up to next-to-leading order in $\alpha_s$ and power corrections up order $1/m_c^5$
        (from Ref.~\cite{Bharucha:2010im}) and next-to-next-to-leading order (NNLO) in $\alpha_s$ (from Ref.~\cite{Grigo:2012ji}).
        The reference values in the $\overline{\text{MS}}$ scheme for the masses of the $c$-quark and $s$-quark
        and the strong coupling are $m_c(m_c) = 1.275\,\GeV$, $m_s(m_c) = 112\,\GeV$,
        and $\alpha_s(m_c) = 0.3996$. For scale-dependent quantities, we use $\mu = m_c(m_c)$.
    }
\end{table}

The various hadronic decay constants and form factors discussed in \cref{app:hme:defs} are genuine nonperturbative
quantities. Nevertheless, perturbation theory can assist in elucidating at least some information about them.
This is achieved with the framework of dispersive bounds; see Ref.~\cite{Caprini:2019osi} for a textbook introduction.
Here, we use a modified formulation of the dispersive bounds~\cite{Gubernari:2023puw}, compared to common formulations~\cite{Okubo:1973tj,Boyd:1997kz,Caprini:1997mu}.
Following Ref.~\cite{Gubernari:2023puw}, we begin with defining a suitable two-point correlation function $\Pi^J_\Gamma(Q^2)$,
\begin{align}
    \label{eq:hme:bounds:pi-tensor}
    \Pi^{\mu\nu}_{\Gamma}(q)
        & \equiv i\! \int\! d^4x\, e^{iq\cdot x} \braket{0 | \mathcal{T} {J_\Gamma^\mu(x) J_\Gamma^{\dagger,\nu}(0)} | 0}
        \,,\\
\intertext{with a decomposition into scalar-valued functions $\Pi^{(i)}$}
    \label{eq:hme:bounds:pi-lambda}
    \Pi^{\mu\nu}_{\Gamma}(q)
        & \equiv \sum_{i} \mathcal{S}^{\mu\nu}_{i} \, \Pi^{(i)}_{\Gamma}(q^2)
        \,.
\end{align}
Here, $\mathcal{S}^{\mu\nu}_{i}$ represents one of two structures with definite angular momentum $J$:
\begin{equation}
    \mathcal{S}^{\mu\nu}_{(J=1)}
        = \left(\frac{q^\mu q^\nu}{q^2} - g^{\mu\nu}\right)\,, \qquad
    \mathcal{S}^{\mu\nu}_{(J=0)}
        = \frac{q^\mu q^\nu}{q^2}
\end{equation}
For our analysis, the relevant currents $J_\Gamma$ are
\begin{equation}
\begin{aligned}
    J_V^\mu(x) & = \bar{s}(x)\, \gamma^\mu c(x)    \,,&\qquad
    J_A^\mu(x) & = \bar{s}(x)\, \gamma^\mu \gamma_5 c(x)      \,,\\
    J_T^\mu(x) & = \bar{s}(x)\, \sigma^{\mu \alpha}q_\alpha  c(x)   \,,&
    J_{AT}^\mu(x) & = \bar{s}(x)\, \sigma^{\mu \alpha}q_\alpha \gamma_5  c(x) \,.
\end{aligned}
\end{equation}
The same currents are also used in the definition of the hadronic decay constants and form factors.
For $Q^2 \lesssim 0$, one finds for the virtuality $Q^2 - (m_c + m_s)^2 \gg \Lambda_\text{had}^2$, ensuring that
$\Pi^{(J)}_\Gamma(Q^2)$ can be computed in a local operator product expansion (OPE).
A number $n = n_\Gamma$ of subtractions,
\begin{align}
    \label{eq:hme:bounds:chisub-lambda}
    \chi_\Gamma^{(i)}(Q^2)
        & = \frac{1}{n!} \left[\frac{\partial}{\partial q^2}\right]^n \Pi_\Gamma^{(i)}(q^2)
        \bigg|_{q^2=Q^2}
          = \frac{1}{\pi} \int\limits_0^\infty ds\, \frac{\Im \Pi_\Gamma^{(i)}(s)}{(s - Q^2)^{n+1}}\,,
\end{align}
is essential to render OPE results for the correlation function finite.
For the vector and axial currents, we use analytic results obtained from a calculation to NNLO
in $\alpha_s$~\cite{Grigo:2012ji} together with analytic results for the contributions by $q\bar{q}$,
$GG$, and $\bar{q}Gq$ vacuum condensates~\cite{Bharucha:2010im}.
For the tensor currents, no NNLO calculation is available. We use the NLO and condensate results obtained
in Ref.~\cite{Bharucha:2010im}.
The values of $\chi_\Gamma^{(i)}$ and the minimal number of subtractions $n$ are compiled
in \cref{tab:hme:chi}.
We find that, although the individual condensate contributions are sizable, they largely cancel when taking their sum.

A hadronic representation of the same correlation functions arises from the computation of the
imaginary part of $\Pi^{(i)}$ in terms of hadronic matrix elements
\begin{equation}
    \label{eq:ImPi}
    \Im\,\Pi_\Gamma^{(i)}(s + i \varepsilon) 
        = \frac{1}{2} \sum \!\!\!\!\!\!\!\! \int\limits_H d\rho_H (2\pi)^4 \delta^{(4)}(p_H - q)
        \mathcal{P}_{\mu\nu}^{(i)}
            \braket{0 | J_\Gamma^\mu | H(q)}\braket{\bar H(q) | J_\Gamma^{\dagger,\nu} \!| 0}\Big|_{q^2=s} 
    \,,
\end{equation}
where $\mathcal{P}_{(J=0)} = \mathcal{S}_{(J=0)}$
and $\mathcal{P}_{(J=1)} = \mathcal{S}_{(J=1)} / 3$.
For one-particle bound states $H = \bar{D}_s^{(*)}$, the contributions read
\begin{equation}
\begin{aligned} 
    \label{eq:1ptcontr}
    &
    \chi^{(J=1)}_{V}\big|_{\mathrm{1pt}} = \frac{M_{D_s^*}^2 f_{D_{s}^*}^2}{(M_{D_s^*}^2-Q^2)^3}\,,  
    &&\quad
    \chi^{(J=0)}_{A}\big|_{\mathrm{1pt}} = \frac{M_{D_s}^2 f_{D_s}^2}{(M_{D_s}^2-Q^2)^2}\,, &
    &&\quad
    \chi^{(J=1)}_{T}\big|_{\mathrm{1pt}} = \frac{M_{D_s^*}^4 (f_{D_s^*}^T)^2}{(M_{D_s^*}^2-Q^2)^4}\,. & 
\end{aligned}
\end{equation}
For $H=D\bar{K}$, the contributions by the hadronic form factors read e.g.~\cite{Gubernari:2023puw}
\begin{align}
\label{eq:chi_time_V}
    \chi^{(J=1)}_{V}\big|_{D\bar{K}}  & = \, \frac{\eta^{D\to K}}{16 \pi^2} \int\limits_{(M_D + M_K)^2}^\infty ds \frac{ \lambda^{3/2}(s)}{s^2 (s - Q^2)^3}
        \, |f_+^{D\to K}(s)|^2 \,, 
\end{align}
where $\lambda(s) \equiv \lambda(M_D^2, M_K^2, s)$ denotes the K\"allen function.
Similar relations exist for all $D\to \bar{K}$ and $\Lambda_c\to \Lambda$ form factors.
The assignment of individual form factors to the quantities $\chi^{(i)}_\Gamma$ is provided in \cref{tab:hme:chi}.
These relations inspire dispersively-bounded parametrisations~\cite{Okubo:1973tj,Boyd:1997kz,Caprini:1997mu}
of the hadronic form factors. We apply this framework in the form discussed in Refs.~\cite{Blake:2022vfl,Amhis:2022vcd,Gubernari:2023puw},
which improves upon previous works by accounting for the integration domain for the dispersive bound and
by splitting the bounds by the helicity as discussed above. The final parametrisation for a form factor $f$ takes the form
\begin{equation}
    \label{eq:hme:ff_param}
    f(q^2) = \frac{1}{\phi_f(z) B(z)} \sum_{k=0}^K a_k^{(f)} p_k^{(f)}(z)\bigg|_{z = z(q^2)}\,,
\end{equation}
where we use the usual conformal map from $q^2$ to $z$, outer functions $\phi_f$, and Blaschke factors $B(z)$.
The functions $p_k^{(f)}$ are a suitable choice of polynomials of order $k$ that are orthonormal on an $f$-specific arc of the unit circle in $z$~\cite{Gubernari:2023puw}.
The manifest benefit of using this parametrisation is the bounded parameter space $|a_k^{(f)}| < 1$ for all orders in $k$
and all form factors $f$. Moreover, in a global fit, we impose a strong dispersive bound of the form
\begin{equation}
    \sum_{f} \sum_{k=0}^K|a_k^{(f)}|^2 < 1\,,
\end{equation}
where the sum over $f$ iterates over all form factors across processes for a fixed current $\Gamma$
and angular momentum $J$, \ie, over all such form factors emerging in either $D\to\bar{K}$ and $\Lambda_c\to\Lambda$.

\subsection{Effective theory relations for baryon form factors}
\label{app:hme:baryonic-ff-relations}

The $\Lambda_c \to \Lambda$ tensor form factors are available from a single lattice QCD analysis~\cite{Meinel:2016dqj}
that provides the (axial)vector and (pseudo)scalar form factors, \ie, all form factors needed to
produce theory predictions within the SM.
For the purpose of our analysis, as discussed in \cref{sec:setup:models}, knowledge of the tensor form factors is also required.
Information on the latter is presently not available from lattice QCD analyses.
In the absence of such lattice QCD information, we rely on other constraints.

Heavy-to-light transition form factors of baryons exhibit some interesting and useful symmetry properties
in the heavy-quark limit (HQL) and in the large-energy limit (LEL), respectively.
These symmetry properties emerge to leading order in the double expansion in $\alpha_s / \pi$ and $\Lambda_\textrm{QCD} / m_c$
and a triple expansion $\alpha_s / \pi$, $\Lambda_\textrm{had} / m_c$, $\Lambda_\textrm{had}/E_{\Lambda}$
(with $E_\Lambda$ the energy of the $\Lambda$ in the $\Lambda_c$ rest frame), respectively.
At leading order in these expansions, all the $\Lambda_c \to \Lambda$ form factors reduce to the following simple set of functions~\cite{Feldmann:2011xf,Mannel:2011xg}
\begin{align}
    \frac{\xi}{m_{\Lambda_c}}
        & = f_{V,t}(0) = f_{V,\perp}(0) = f_{V,0}(0) = f_{A,t}(0) = f_{A,\perp}(0) = f_{A,0}(0) \nonumber \\
        & = f_{T,\perp}(0) = f_{T,0}(0) = f_{T5,\perp}(0) = f_{T5,0}(0) \,, \\
    \frac{\xi_1 - \xi_2}{m_{\Lambda_c}}
        & = f_{V,\perp}(q^2_\mathrm{max}) = f_{V,0}(q^2_\mathrm{max}) = f_{A,t}(q^2_\mathrm{max}) =
            f_{T,\perp}(q^2_\mathrm{max}) = f_{T,0}(q^2_\mathrm{max}) \,, \\
    \frac{\xi_1 + \xi_2}{m_{\Lambda_c}}
        & = f_{A,\perp}(q^2_\mathrm{max}) = f_{A,0}(q^2_\mathrm{max}) = f_{V,t}(q^2_\mathrm{max}) =
            f_{T5,\perp}(q^2_\mathrm{max}) = f_{T5,0}(q^2_\mathrm{max}) \,. 
\end{align}
Although the values of $\xi, \xi_1$ and $\xi_2$ can be approximated, a proper estimation of the correlation between the different form factors would require dedicated analyses.
Here, we instead follow the recipe proposed in Ref.~\cite{Amhis:2022vcd} and restrict ourselves to the following set of relations,
valid both at $q^2 = q^2_\mathrm{max}$ (HQL) and $q^2 = 0$ (LEL):
\begin{equation}
\begin{aligned}
    f_{T,\perp} / f_{V,\perp}  &=\, 1 \pm 0.35 \,, &  f_{T,0}  / f_{V,0} &=\, 1 \pm 0.35 \,, \\
    f_{T5,\perp} / f_{A,\perp} &=\, 1 \pm 0.35 \,, &  f_{T5,0} / f_{A,0} &=\, 1 \pm 0.35 \,,
\end{aligned}
\end{equation} 
where the uncertainties are treated as uncorrelated.
We find that imposing these relations suffices to determine the tensor form factor parameters in the presence of the dispersive bound.

\bibliographystyle{jhep} 
\bibliography{references.bib}

\end{document}